\documentclass[aps,prl,reprint,lengthcheck,superscriptaddress,showpacs]{revtex4-1}
\pdfoutput=1

\usepackage{atlasphysics}
\usepackage{multirow}
\usepackage{graphicx}
\usepackage{color}
\usepackage{hyperref}
\usepackage{longtable}
\usepackage{dcolumn}
\usepackage{comment}
\usepackage{amsmath}

\DeclareGraphicsExtensions{.pdf}

\newcommand{\alphalp}{\ensuremath{\alpha_\ell P}}
\newcommand{\alphaonep}{\ensuremath{\alpha_1 P_1}}
\newcommand{\alphatwop}{\ensuremath{\alpha_2 P_2}}

\newcommand{\alphalpcpc}{\ensuremath{\alpha_\ell P_{\mathrm{CPC}}}}
\newcommand{\alphalpcpv}{\ensuremath{\alpha_\ell P_{\mathrm{CPV}}}}
\newcommand{\ljets}{$\ell+$jets}
\newcommand{\ejets}{$e+$jets}
\newcommand{\mujets}{$\mu+$jets}
\newcommand{\emu}{$e\mu$}
\newcommand{\mm}{$\mu\mu$}
\newcommand{\elel}{$ee$}

\usepackage{preprintcover} 
\PreprintCoverPaperTitle{\protect\boldmath Measurement of top quark polarization in top--antitop events from proton--proton collisions at $\boldsymbol{\sqrt{s}=7}$~\TeV\ using the ATLAS detector} 
\PreprintIdNumber{CERN-PH-EP-2013-101} 
\PreprintCoverAbstract{This Letter presents measurements of the polarization of the top quark in top--antitop quark pair events, using  4.7~\ifb\ of proton--proton collision data recorded with the ATLAS detector at the Large Hadron Collider at $\sqrt{s}=7~\TeV$. 
Final states containing one or two isolated leptons (electrons or muons) and jets are considered.  Two measurements of \alphalp, the product of the leptonic spin-analyzing power and the top quark polarization, are performed assuming that the polarization is introduced by either a CP conserving (CPC) or a CP violating (CPV) production process. 
The measurements obtained, $\alphalpcpc = -0.035 \pm 0.014\mathrm{(stat)} \pm 0.037 \mathrm{(syst)}$ and $\alphalpcpv = 0.020 \pm 0.016\mathrm{(stat)}^{+0.013}_{-0.017} \mathrm{(syst)}$, are in good agreement with the Standard Model prediction of negligible top quark polarization.} 
\PreprintJournalName{PRL}

\begin{document}

\title{Measurement of top quark polarization in top--antitop events from proton--proton collisions at $\boldsymbol{\sqrt{s}=7}$~\TeV\ using the ATLAS detector}
\collaboration{The ATLAS Collaboration}

\date{September 9, 2013}

\begin{abstract}
This Letter presents measurements of the polarization of the top quark in top--antitop quark pair events, using  4.7~\ifb\ of proton--proton collision data recorded with the ATLAS detector at the Large Hadron Collider at $\sqrt{s}=7~\TeV$. 
Final states containing one or two isolated leptons (electrons or muons) and jets are considered.  Two measurements of \alphalp, the product of the leptonic spin-analyzing power and the top quark polarization, are performed assuming that the polarization is introduced by either a CP conserving (CPC) or a CP violating (CPV) production process. 
The measurements obtained, $\alphalpcpc = -0.035 \pm 0.014\mathrm{(stat)} \pm 0.037 \mathrm{(syst)}$ and $\alphalpcpv = 0.020 \pm 0.016\mathrm{(stat)}^{+0.013}_{-0.017} \mathrm{(syst)}$, are in good agreement with the Standard Model prediction of negligible top quark polarization.
\end{abstract}


\maketitle

The short lifetime of the top
quark~\cite{Aaltonen:2010ea,Aaltonen:2008ir,Abazov:2010tm,Abazov:2012vd,Jezabek:1993wk}
implies that it decays before hadronization takes place, allowing its spin state to be studied using the angular distributions of its decay
products.  In the Standard Model (SM), parity conservation in the
strong production of top--antitop quark pairs (\ttbar) in proton--proton ($pp$) collisions implies zero longitudinal polarization of the quarks.  A negligible polarization (0.003) is generated by the weak interaction~\cite{Bernreuther:2013aga}.  Physics beyond the SM can
induce top quark polarization.  For example, models that predict the
top quark forward-backward production asymmetry to be larger than the SM
prediction, as seen by the Tevatron experiments D0~\cite{d0asymmetry,d0asymmetrydilep} and
CDF~\cite{cdfasymmetry}, can generate
non-zero polarization of top quarks~\cite{Krohn:2011tw,Fajfer:2012si,aguilar2013}. A
first study of polarization in \ttbar\ events has been performed by
the D0 collaboration~\cite{d0asymmetrydilep}, showing good agreement between the SM
prediction and data.

In this Letter, measurements are presented of the polarization of the top quark in inclusive \ttbar\ production in single charged lepton ($\ttbar\rightarrow \ell\nu q\bar{q} b\bar{b}$) and dilepton ($\ttbar\rightarrow \ell^+\nu\ell^-\bar{\nu} b\bar{b}$) events.  The double differential distribution in polar angles, $\theta$, of two of the final-state decay products, with respect to a given quantization axis is given by~\cite{bernreutherpol}
\begin{eqnarray}
 \frac{1}{\sigma} \frac{d\sigma}{d\cos\theta_1 d\cos\theta_2} & =&
 \frac{1}{4}\left( 1 + \alphaonep\ \cos\theta_1 + \alphatwop\ \cos\theta_2 \right. \nonumber \\ 
 & & \left. {}  - C \cos\theta_1  \cos\theta_2  \right), 
\label{eq:coscos}
\end{eqnarray}
where $\theta_1$ ($\theta_2$) is the angular distribution of the decay daughter particle of the top (antitop) quark. Here, $C$ represents the $t\bar{t}$ spin correlation, $P_1$ ($P_2$) represents the degree of polarization of the top (antitop) quark along the chosen quantization axis, and $\alpha_i$ is the spin-analyzing power of the final state object~\cite{SpinPower,Mahlon:1995zn}, which is a measure of the sensitivity of the daughter particle to the spin state of the parent.  At leading order, charged leptons and down-type quarks from $W$-boson decays are predicted to have the largest sensitivity to the spin state of the top quark with a spin-analyzing power of ${\alpha=1}$. The helicity basis is used, in which the momentum direction of the top quark in the \ttbar\ center-of-mass frame is chosen as the  quantization axis.  The $\cos\theta_\ell$ distributions of the charged leptons are used as observables to extract a measurement of \alphalp.

The analysis is based on the full 2011 dataset of $pp$ collision events, collected at a center-of-mass energy of 7~\TeV\ by the ATLAS detector~\cite{myATL-2008-001}, corresponding to an integrated luminosity of 4.66 $\pm$ 0.08 \ifb~\cite{lumi} after data quality requirements. 

ATLAS includes an inner tracking detector, covering a pseudorapidity~\cite{geometry} range ${|\eta|<2.5}$,  surrounded by a superconducting solenoid providing a 2 T magnetic field. A liquid argon (LAr) electromagnetic sampling calorimeter (${|\eta|<3.2}$), an iron--scintillator tile hadronic calorimeter (${|\eta|<1.7}$), a LAr hadronic calorimeter (${1.4<|\eta|<3.2}$), and a LAr forward calorimeter (${3.1<|\eta|<4.9}$) provide the energy measurements. The muon spectrometer consists of tracking chambers covering ${|\eta|<2.7}$, and trigger chambers covering ${|\eta|<2.4}$, in a toroidal magnetic field.  Events considered in this analysis are required to have one high-transverse-momentum (\pt) electron or muon that passes requirements of the three-level trigger system.

Both data-driven techniques and Monte Carlo (MC) simulations are used to estimate the sample composition of the data. For each MC sample, generated events are processed through a \textsc{Geant4}~\cite{myAGO-0301} simulation of the full ATLAS detector~\cite{atlasSimulation}, and the same reconstruction and analysis software is used for both the data and the MC events.  Signal \ttbar\ events are simulated by the next-to-leading-order (NLO) generator \textsc{MC@NLO 3.41}~\cite{myFRI-0201} with the NLO parton distribution function (PDF) set CT10~\cite{Lai}, assuming a top quark mass of 172.5 \GeV.  Parton showering is modeled with \textsc{HERWIG 6.510}~\cite{myCOR-0001}, and \textsc{JIMMY 4.31}~\cite{myJimmy} is used for the underlying event. A \ttbar\ production cross section of $167^{+ 17}_{- 18}$ pb is used, calculated at approximate next-to-next-to-leading-order (NNLO) in QCD using \textsc{Hathor} 1.2~\cite{hathorxs}.
Backgrounds are simulated using the \textsc{MC@NLO}, \textsc{AcerMC}~\cite{SAMPLES-ACER}, \textsc{ALPGEN}~\cite{myMAN-0301}, and \textsc{HERWIG} generators, as detailed in Ref.~\cite{Aad:2013nca}. Each simulated signal or background event is overlaid with additional $pp$ collisions.  The events are given a weight such that the distribution of the average number of events per beam crossing agrees with data.  For each sample the cross section is rescaled to the most up to date theoretical expectations, as described in Ref.~\cite{Aad:2010ey}. 

The data sample is enriched in \ttbar\ events by applying several selection criteria based on the \ttbar\ event topology. The selected \ttbar\ events consist of jets, isolated leptons and missing transverse momentum from the undetected neutrinos. 
Jets are reconstructed from clustered energy deposits in the calorimeters using the anti-$k_t$ algorithm~\cite{antikt} with a radius parameter ${R = 0.4}$. Their energies are corrected  to correspond on
average to the total energy of the stable particles emitted towards
the jet using energy- and $\eta$-dependent correction factors derived from simulation, and a residual correction derived from in situ measurements~\cite{Aad:2010ad,ATLAS-CONF-2013-004}. They are required to have ${\pt > 25 \GeV}$ and  ${|\eta|<2.5}$.  Furthermore, at least 75\% of the scalar sum of the \pt\ of all the tracks associated with each jet must belong to tracks originating from the primary vertex, which is defined as the vertex with the highest sum of the squared \pt\ values of the associated tracks in the event. 
Jets originating from $b$-quarks are selected by using a neural network algorithm that combines information about the impact parameter significance of tracks with information about explicitly reconstructed secondary vertices and other variables. At the chosen working point, the algorithm identifies simulated $b$-jets from top quark decays with 70\% efficiency and a rejection factor of about 140 for light partons~\cite{ATLAS-CONF-2011-102,ATLAS-CONF-2012-043,ATLAS-CONF-2012-040}.
Reconstructed electrons must have ${\pt\ > 25}$~\GeV\ and be associated with a calorimeter cluster in the range ${|\eta_{\mathrm{cl}}| < 2.47}$, excluding the transition between calorimeter sections, ${1.37<|\eta_{\mathrm{cl}}| <1.52}$.  Selected muons are required to fulfill ${|\eta| < 2.5}$ and  ${\pt > 20 \GeV}$.  Each lepton is required to pass quality criteria, to be compatible with being produced at the primary vertex by having a longitudinal impact parameter smaller than 2 mm, and to be isolated from other calorimeter energy deposits and tracks~\cite{Aad:2013rna}.
The \met\ is calculated~\cite{EtMiss} as the magnitude of the negative of the vectorial sum of all energy deposits in the calorimeters, and then corrected for the momenta of the reconstructed muons. 

The details of the final event selection depend on the \Wboson\ decay channels. This measurement uses five different channels, containing either one or two electrons or muons in the final state, including the ones coming from $\tau$ decays. The requirements for the single-lepton channels (\ljets) are: 
\begin{itemize}
\item Exactly one electron or muon;
\item At least four jets, at least one of which is $b$-tagged;
\item ${\met > 30 \GeV}$ for the electron channel and ${\met > 20 \GeV}$ for the muon channel;
\item The transverse mass of the \Wboson\ boson to be greater than $30 \GeV$  for the electron channel, while ${m_{\mathrm{T}} + \met > 60 \GeV}$ is required for the muon channel. The transverse mass is computed from the lepton \pt\ and $\phi$ angle ($\pt^\ell$, $\phi^\ell$) and the direction of the \met\ as $m_\mathrm{T}= \sqrt{2\pt^\ell \met [1-\cos(\phi^\ell - \phi(\met))]}$.
\end{itemize}
The selection of the dilepton channels (\elel, \emu, \mm) requires:
\begin{itemize}
  \item Exactly two oppositely charged electrons or muons;
  \item At least two jets;
  \item A dilepton invariant mass larger than 15~\GeV\ for all the channels, and more than 10~\GeV\ away from the $Z$
    boson mass for the $ee$ and $\mu\mu$ channels;
  \item ${\met > 60\GeV}$ for the $ee$ and $\mu\mu$ channels;
  \item The scalar sum of the \pt\ of all selected leptons and jets to be larger than 130~\GeV\ for the $e\mu$ channel. 
\end{itemize}

The major backgrounds are due to vector boson production with additional jets, single top quark production, and to misidentified leptons.  Their contributions are estimated using data-driven methods and MC simulation. In particular, the normalization of the dominant background in the \ljets\ channels,  $W+$jets production, is estimated using a measurement of the lepton charge asymmetry in data~\cite{ATLAS:2012an}, while the shape of the distribution of $\cos\theta_\ell$ is taken from simulation.  
In the \elel\  (\mm) channel, the normalization of the $Z/\gamma^{*}+$jets background with $Z/\gamma^{*}$ decaying into \elel\ (\mm) is determined from data. A $Z/\gamma^{*}+$jets enriched control region is defined, where a correction factor for the simulation normalization is derived as a function of the \met\ in the event, and applied to the signal region in order to account for possible \met\ mis-modeling. 

The contributions of non-prompt leptons from semi-leptonic hadron decays and of jets misidentified as leptons (fakes) are determined from data using matrix methods~\cite{Aad:2010ey,Aad:2012hg}.  For \ljets\ channels this contribution comes primarily from multi-jet events, while for dilepton channels it originates primarily from $W+$jets events where one charged lepton comes from $W$ decay and the other lepton is a non-prompt or fake lepton.

After selection, the expected yields for signal and background compared to data are shown in Table~\ref{tab:yields}.
\begin{table}
\caption{Expected signal and background rounded yields compared to data for each of the five lepton flavor channels considered. The approximate NNLO SM prediction~\cite{hathorxs} is assumed for $t\bar{t}$ production, and the total systematic and statistical uncertainties are reported.}
\begin{ruledtabular}
\begin{tabular}{crrrrr}
Source  & \ejets & \mujets & \elel & \emu & \mm \\ \hline \\[-6pt]
\ttbar  & 16200  & 26500   & 570   & 4400 & 1660 \\ 
Background   & 5100   & 9400    & 110   & 700  & 320 \\[10pt]
Total   & 21300  & 35900   & 690   & 5000 & 1980 \\ 
Uncertainty & $\pm1300$ & $\pm1700$ & $\pm80$ & $\pm500$ & $\pm180$ \\[10pt]
Data    & 21956  & 37919   & 740   & 5328 & 2057 \\
\end{tabular}
\end{ruledtabular}
\label{tab:yields}
\end{table}

The selected events are reconstructed under the \ttbar\ event
hypothesis: jets are associated with particular quarks, and the longitudinal momenta of the neutrinos in the event are
determined. From the fully reconstructed decay chain we calculate the
momentum of the top quark in the \ttbar\ frame and from it
$\cos\theta_l$.

In the \ljets\ channels, a kinematic likelihood fit is performed.  The likelihood for the event to correspond to a \ttbar\ decay topology is calculated for each possible assignment of four jets selected from the up to five highest \pt\ jets in the event, to the two $b$-quark jets and the two jets from the $W$-boson decay~\cite{ATLAS:2012aj}.  The energies of the jets and the charged lepton, as well as the \met, are allowed to vary within their respective resolutions to best meet the $W$-boson and top quark mass constraints to form the kinematic likelihood.  The combined probability is calculated as the product of the maximum kinematic likelihood, the $b$-tagging efficiency and light-parton rejection probability.  The highest probability permutation is chosen as the best reconstruction and used to calculate the charged lepton $\cos\theta_\ell$.

In the dilepton channels, the neutrino weighting method is used~\cite{nuw}. Because of the presence of two neutrinos from $W$ boson decays, the final system is underconstrained and assumptions must be made to calculate all particle momenta.  Making a hypothesis for the pseudorapidities of the two neutrinos ($\eta_1,\eta_2$), a weight is assigned for each permutation of jets,  based on the compatibility of the total neutrino transverse momentum vector and the measured \met, accounting for \met\ resolution~\cite{EtMiss}. For each event, 10000 different hypotheses for ${(\eta_1,\,\eta_2)}$ are scanned, drawn from the observed probability distribution in the signal MC sample.  The configuration with the maximal weight is selected and used to reconstruct the values of $\cos\theta_\ell$ for both charged leptons. Events for which no physical solution can be found with this method are discarded, corresponding to 15\% of the selected events in the simulated dilepton \ttbar\ sample.  The assumed $\eta$ distributions of the neutrinos are insensitive to top quark polarization.

To extract the value of \alphalp\ from the data, a fit using templates for partially polarized top quarks is performed. The signal templates are obtained by reweighting the top and antitop quark decay products in the simulated \ttbar\ sample according to Eq.~\ref{eq:coscos} using the helicity basis and setting $C$ to the SM value of $t\bar{t}$ spin correlation, $C=0.31$~\cite{Bernreuther:2013aga}.  Two different assumptions about the top quark polarization are made to produce two template fits.  In one case, the polarization is assumed to be induced by a charge--parity (CP) conserving process, which leads to top and antitop quarks having equal values of \alphalp\ and therefore the same angular distribution for the daughter particles.  In the other, maximal CP violation is assumed, leading to opposite values of \alphalp\ for the top and antitop quarks.  In this case, when a value of \alphalp\ is quoted its sign refers to the sign of the coefficient for positively charged leptons.  The positive and negative templates used in the fit are built assuming a value of ${\alphalp\ = \pm 0.3}$, to guarantee that the differential decay distribution is positive for all values of $\cos\theta_\ell$ given the degree of spin correlation. The fraction, $f$, of the positive template component and the \ttbar\ production cross section are fitted simultaneously, in order to reduce the influence of normalization uncertainties on the measured polarization. The polarization is computed as ${\alphalp  = 0.6 f - 0.3}$.

For all the considered channels, a template fit is performed with a binned maximum likelihood method on positive and negative lepton distributions separately.
Combinations are made by multiplying the likelihood functions of the single channels for the two single-lepton channels, the dilepton channels, and all channels together.
The fitting method is unbiased, which was shown using pseudo-experiments.  

For each source of systematic uncertainty,  new templates corresponding to the respective one standard deviation up and down variation are considered. When an uncertainty is evaluated as the difference between two points, it is symmetrized around the central  value. 
The mean of the distribution of the respective differences between the central fit values and the up and down results from 1000
pseudo-datasets are taken as the systematic uncertainties on that source.  Systematic uncertainties arising from the same source are treated as being correlated between the different lepton charge and flavor samples.

Detector systematic uncertainties, related to the determination of the energy or momentum scales, resolutions, and efficiencies for jets, electrons, and muons, as well as the \met\ are considered~\cite{JetScalePaper,ATLAS-CONF-2013-004,JetResPaper,Aad:2011mk,MuonEffConf,MuonResConf,EtMiss}. Simulated samples are corrected in order to match the reconstructed object properties observed in data, and the correction factors are varied depending on the uncertainties of their values, in order to estimate the uncertainty on the final measurement. The largest uncertainty in this measurement comes from the jet energy scale. 

Systematic uncertainties from the modeling of the \ttbar\ final state in simulation are accounted for using alternative signal templates. These templates are produced by varying the MC event generator, initial- and final- state radiation, color reconnection, fragmentation modeling, and the PDF sets, as detailed in Ref.~\cite{ttbarResonanceATLAS}. The estimation of the uncertainty due to the top quark mass is performed by repeating the fitting procedure using seven samples with different mass settings in the simulation,  and interpolating the change in the parameter $f$ corresponding to a variation of the top quark mass of $\pm 1.4$ \GeV~\cite{ATLAS:2012coa} around the nominal value.  Because an assumption on the degree of spin correlation is made when constructing the template, an additional uncertainty is applied based on the difference in the parton-level spin correlation in simulated \ttbar\ events between the \textsc{MC@NLO} and \textsc{POWHEG}~\cite{myFRI-0701} generators.

For the $W+$jets background in the \ljets\ final state, the overall normalization is varied according to the residual uncertainty after the rescaling based on the measured charge asymmetry~\cite{ATLAS:2012an}.  In addition, the $W+$jets template is varied in shape and normalization by reweighting events according to both the uncertainty in the associated heavy quark production flavor fractions and the parameters of the simulation of extra jets~\cite{Aad:2012hg}.  For the estimate of the systematic uncertainty due to events with non-prompt or fake leptons, the templates are varied according to its uncertainties in the matrix method inputs~\cite{Aad:2010ey,Aad:2012hg}.
The MC statistical uncertainty is taken into account by performing pseudo-experiments, where the bin content of each template is varied independently according to the uncertainty. 
Table~\ref{tab:systs} summarizes the sources of systematic uncertainty and their effect on \alphalp\ for the combined fit. The two largest uncertainties come from jet reconstruction and MC modeling, both affecting the shape of the $\cos\theta_\ell$ distribution. For sources of systematic uncertainty that do not depend on the lepton charge in the event, the uncertainty in the CP violating scenario is greatly reduced.  These uncertainties push the fit parameters in opposite directions for the samples with different lepton charge, leading to a smaller total uncertainty in the combination.

\begin{table}
\caption{Summary of the systematic uncertainties on \alphalp\ for the CP conserving and CP violating fits in the combined channels.  The systematic uncertainties have been added in quadrature to obtain the total uncertainty.}
\begin{ruledtabular}
\begin{tabular}{cccccc}
Source & \multicolumn{2}{c}{$\Delta \alphalpcpc$} & & \multicolumn{2}{c}{$\Delta \alphalpcpv$} \\ \hline
Jet reconstruction  & $+0.031$ & $-0.031$  & & $+0.009$ & $-0.005$ \\ 
Lepton reconstruction  & $+0.006$ & $-0.007$  & & $+0.002$ & $-0.001$ \\ 
\met\ reconstruction  & $+0.008$ & $-0.007$ &   & $+0.004$ & $-0.001$ \\ 
\ttbar\ modeling  & $+0.015$ & $-0.016$  & & $+0.005$ & $-0.013$ \\ 
Background modeling  & $+0.011$ & $-0.010$  & & $+0.005$ & $-0.007$ \\ 
Template statistics  & $+0.005$ & $-0.005$  & & $+0.006$ & $-0.006$ \\[10pt] 
Total systematic uncertainty & $+0.037$ & $-0.037$  & & $+0.013$ & $-0.017$ \\ 
\end{tabular}
\end{ruledtabular}
\label{tab:systs}
\end{table}

The results of the fit to the data in single-lepton and dilepton channels are summarized in Table~\ref{tab:channels}.  Figure~\ref{fig:fit_ljets} shows the fitted observable in the single-lepton and dilepton final states with the CP conserving hypothesis, and Fig.~\ref{fig:fit_ljets_CPV} shows the same observable in the CP violating hypothesis. The deviation from the expected linear behavior of the $\cos\theta_\ell$ distributions is primarily a result of the detector acceptance.

\begin{table}
\caption{Summary of fitted \alphalp\ in the individual channels for the CP conserving and CP violating fits.  The uncertainties quoted are first statistical and then systematic.}
\begin{ruledtabular}
\begin{tabular}{ccc}
Channel & \alphalpcpc & \alphalpcpv \\ \hline
\elel & $ 0.12 \pm 0.10 {}^{ +0.09 }_{ -0.12 } $ & $ -0.04 \pm 0.12 {}^{ +0.18 }_{ -0.12 } $ \\[2pt] 
\emu & $ -0.07 \pm 0.04 {}^{ +0.05 }_{ -0.06 } $ & $ 0.00 \pm 0.04 {}^{ +0.05 }_{ -0.04 } $ \\[2pt]  
\mm & $ -0.04 \pm 0.06 {}^{ +0.07 }_{ -0.07 } $ & $ 0.04 \pm 0.07 {}^{ +0.06 }_{ -0.06 } $ \\[10pt]
Dilepton & $ -0.04 \pm 0.03 {}^{ +0.05 }_{ -0.05 } $ & $ 0.01 \pm 0.03 {}^{ +0.04 }_{ -0.04 } $ \\[10pt]
\ejets & $ -0.031 \pm 0.028 {}^{ +0.043 }_{ -0.040 } $ & $ 0.001 \pm 0.031 {}^{ +0.019 }_{ -0.019 } $ \\[2pt]  
\mujets & $ -0.033 \pm 0.021 {}^{ +0.039 }_{ -0.039 } $ & $ 0.036 \pm 0.023 {}^{ +0.018 }_{ -0.017 } $ \\[10pt]
\ljets & $ -0.034 \pm 0.017 {}^{ +0.038 }_{ -0.037 } $ & $ 0.023 \pm 0.019 {}^{ +0.012 }_{ -0.011 } $ \\[10pt]
Combined & $ -0.035 \pm 0.014 {}^{ +0.037 }_{ -0.037 } $ & $ 0.020 \pm 0.016 {}^{ +0.013 }_{ -0.017 } $ \\  
\end{tabular}
\end{ruledtabular}
\label{tab:channels}
\end{table}

\begin{figure}[h]
\begin{center}
\begin{minipage}{9.0 cm}
\setlength{\unitlength}{1.0cm}
\begin{picture}(9.0,9.0)
\put(-0.3,4.3){\includegraphics[width=\columnwidth]{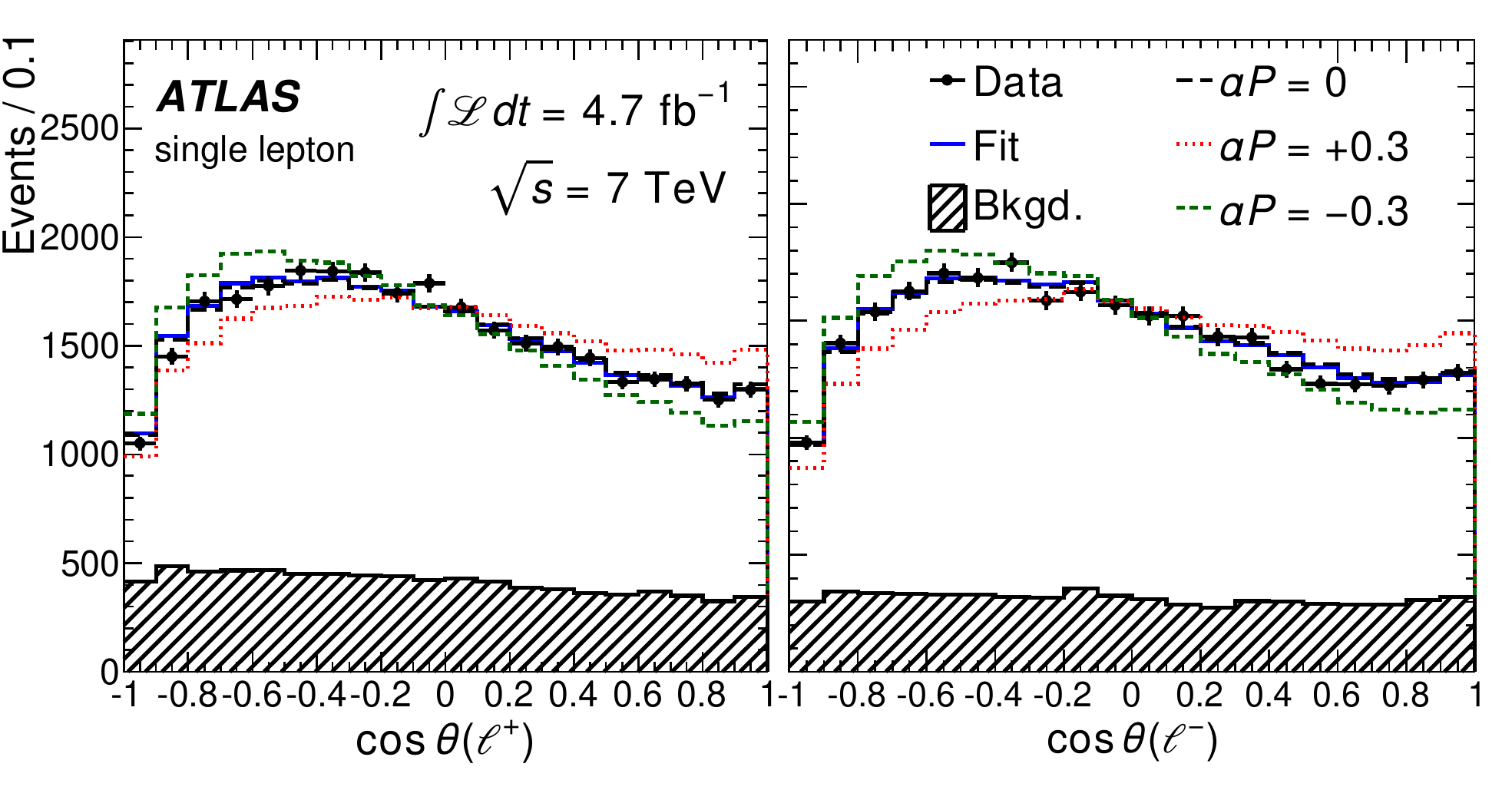}}
\put(-0.3,-0.5){\includegraphics[width=\columnwidth]{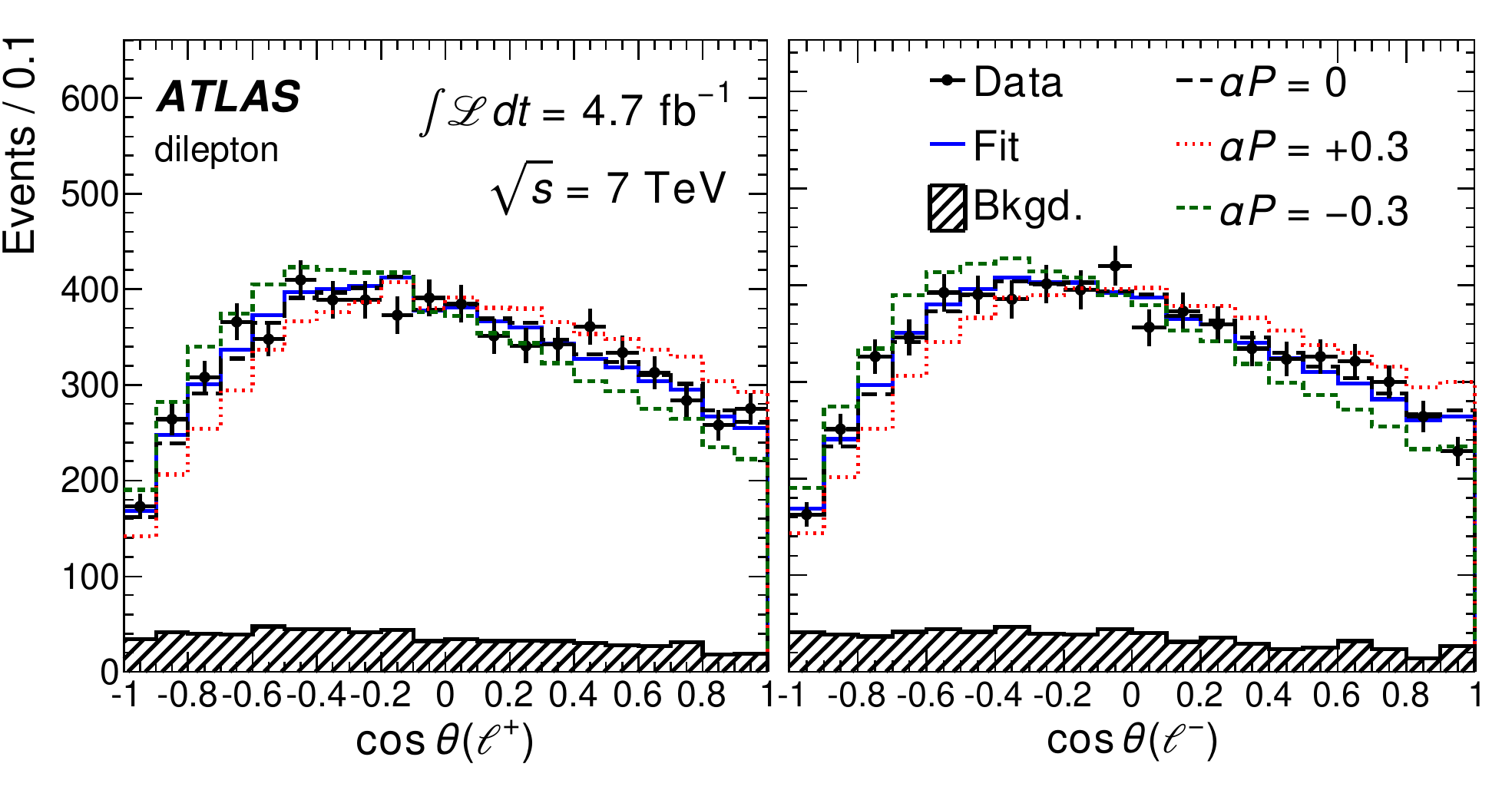}}
\put(1.5,1.5){(b)}
\put(1.5,6.5){(a)}
\end{picture}
\end{minipage}
\end{center}
\caption{The result of the full combined fit to the data with the CP conserving polarization hypothesis in (a) the single-lepton channel and (b) the dilepton channel, adding together electrons and muons. It is compared to the polarization templates used and the SM prediction of zero polarization.  Positively charged leptons are on the left, and negatively charged leptons on the right.}
\label{fig:fit_ljets}
\end{figure}

\begin{figure}[h]
\begin{center}
\begin{minipage}{9.0 cm}
\setlength{\unitlength}{1.0cm}
\begin{picture}(9.0,9.0)
\put(-0.3,4.3){\includegraphics[width=\columnwidth]{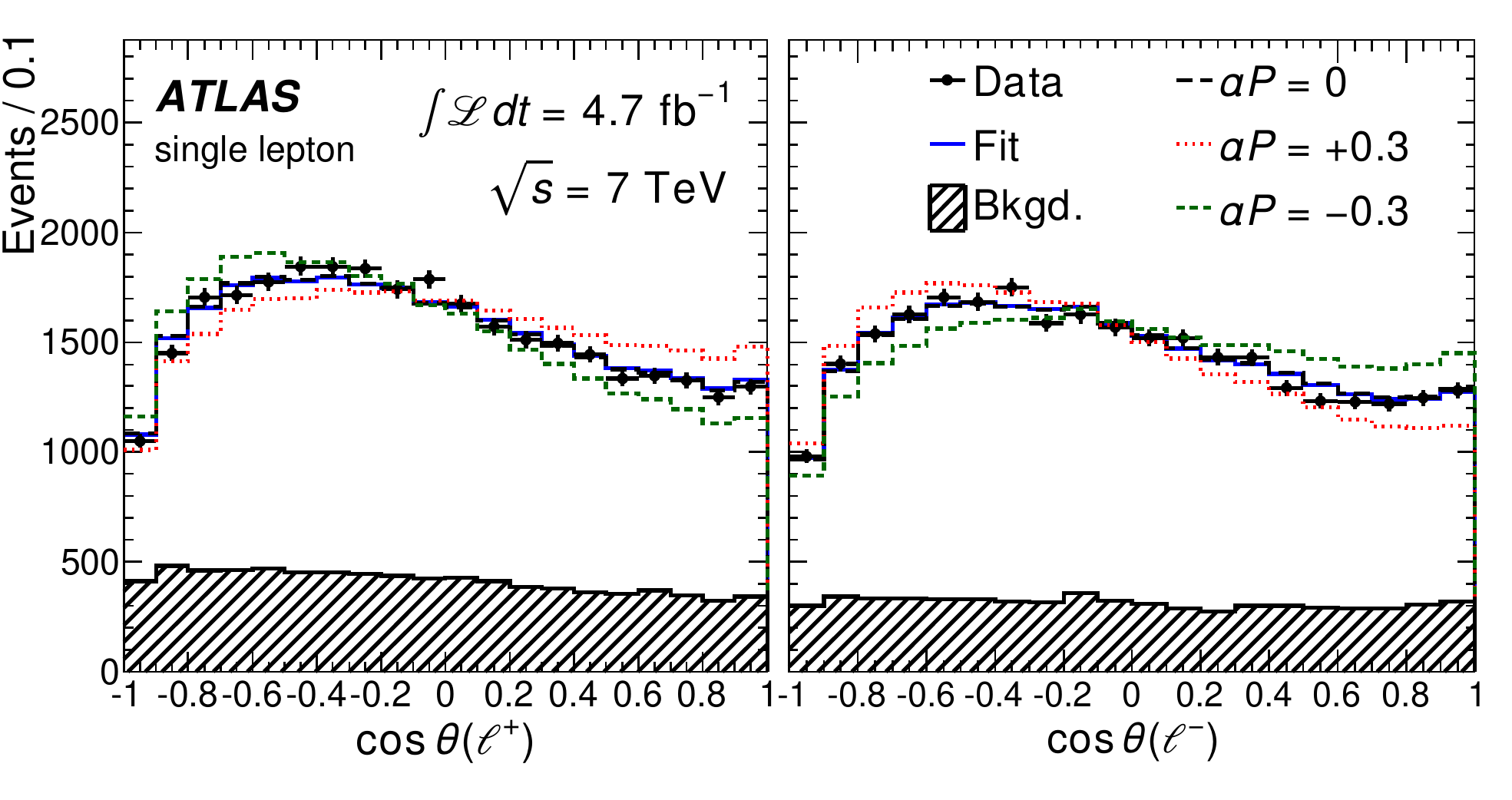}}
\put(-0.3,-0.5){\includegraphics[width=\columnwidth]{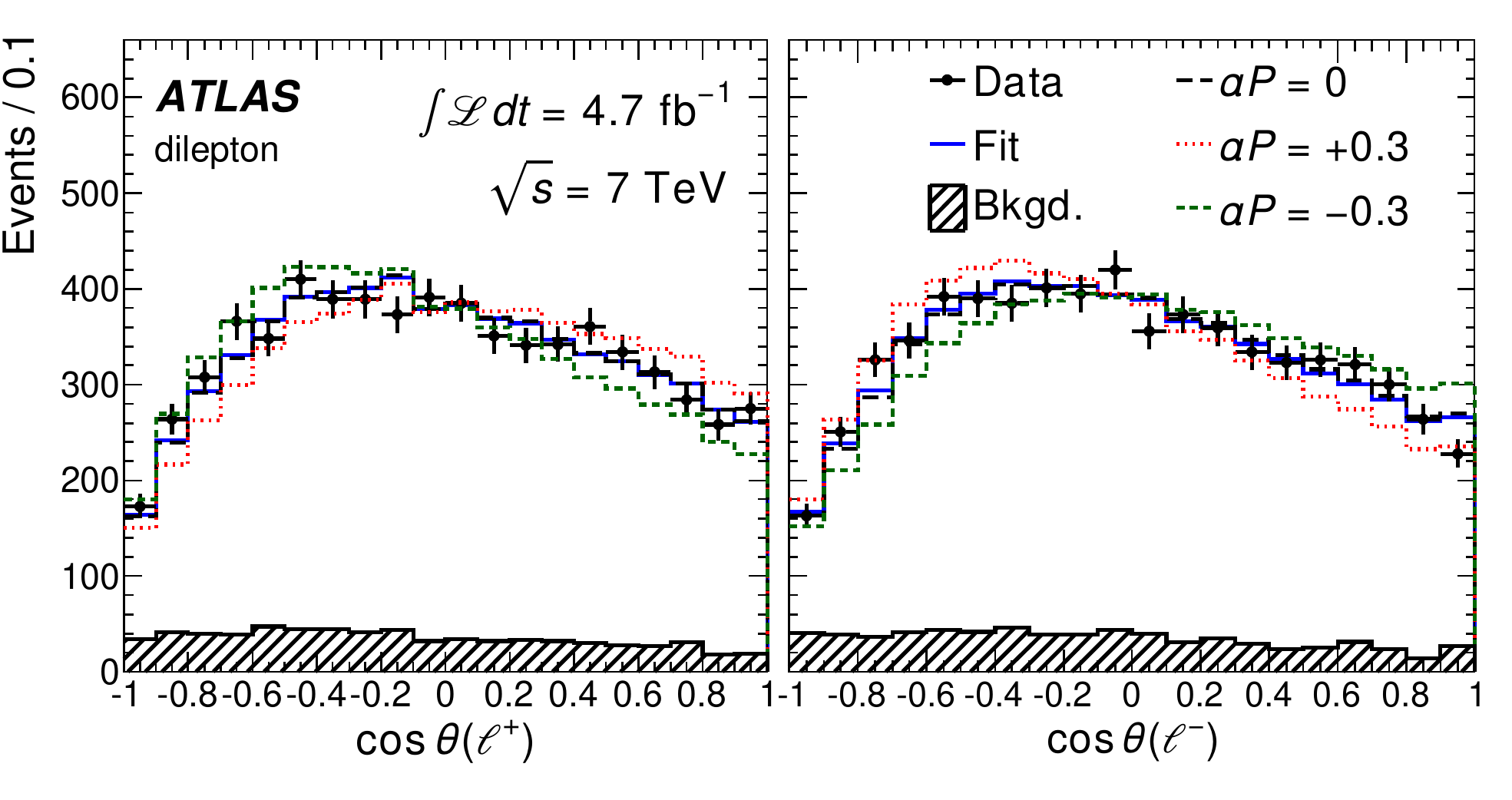}}
\put(1.5,1.5){(b)}
\put(1.5,6.5){(a)}
\end{picture}
\end{minipage}
\end{center}
\caption{The result of the full combined fit to the data with the CP violating polarization hypothesis in (a) the single-lepton channel and (b) the dilepton channel, adding together electrons and muons. It is compared to the polarization templates used and the SM prediction of no polarization.  Positively charged leptons are on the left, and negatively charged leptons on the right.}
\label{fig:fit_ljets_CPV}
\end{figure}

The single-lepton and dilepton channels combined results are: 
\begin{equation}
\alphalpcpc  = -0.035 \pm 0.014\mathrm{(stat)} \pm 0.037\mathrm{(syst)}
\end{equation}
in the CP conserving scenario, and 
\begin{equation}
\alphalpcpv  = 0.020 \pm 0.016\mathrm{(stat)}^{+0.013}_{-0.017} \mathrm{(syst)}
\end{equation}
in the CP violating scenario. The polarization in both scenarios agrees with the SM prediction of negligible polarization. The fitted $\sigma_{\ttbar}$ is in good agreement with the SM prediction as obtained from NNLO QCD calculations~\cite{Czakon:2013goa,Czakon:2011xx}. 

In conclusion, the first measurement of top quark polarization in $t\bar{t}$ events has been performed for two different scenarios with 4.7~\ifb\ of proton--proton collision data at 7~\TeV\ center-of-mass energy with the ATLAS detector at the LHC. Single-lepton and dilepton final states have been used and no deviation from the SM prediction of negligible polarization is observed for either the CP conserving or CP violating scenario.

\begin{acknowledgments}
We thank CERN for the very successful operation of the LHC, as well as the
support staff from our institutions without whom ATLAS could not be
operated efficiently.

We acknowledge the support of ANPCyT, Argentina; YerPhI, Armenia; ARC,
Australia; BMWF and FWF, Austria; ANAS, Azerbaijan; SSTC, Belarus; CNPq and FAPESP,
Brazil; NSERC, NRC and CFI, Canada; CERN; CONICYT, Chile; CAS, MOST and NSFC,
China; COLCIENCIAS, Colombia; MSMT CR, MPO CR and VSC CR, Czech Republic;
DNRF, DNSRC and Lundbeck Foundation, Denmark; EPLANET, ERC and NSRF, European Union;
IN2P3-CNRS, CEA-DSM/IRFU, France; GNSF, Georgia; BMBF, DFG, HGF, MPG and AvH
Foundation, Germany; GSRT and NSRF, Greece; ISF, MINERVA, GIF, DIP and Benoziyo Center,
Israel; INFN, Italy; MEXT and JSPS, Japan; CNRST, Morocco; FOM and NWO,
Netherlands; BRF and RCN, Norway; MNiSW, Poland; GRICES and FCT, Portugal; MERYS
(MECTS), Romania; MES of Russia and ROSATOM, Russian Federation; JINR; MSTD,
Serbia; MSSR, Slovakia; ARRS and MIZ\v{S}, Slovenia; DST/NRF, South Africa;
MICINN, Spain; SRC and Wallenberg Foundation, Sweden; SER, SNSF and Cantons of
Bern and Geneva, Switzerland; NSC, Taiwan; TAEK, Turkey; STFC, the Royal
Society and Leverhulme Trust, United Kingdom; DOE and NSF, United States of
America.

The crucial computing support from all WLCG partners is acknowledged
gratefully, in particular from CERN and the ATLAS Tier-1 facilities at
TRIUMF (Canada), NDGF (Denmark, Norway, Sweden), CC-IN2P3 (France),
KIT/GridKA (Germany), INFN-CNAF (Italy), NL-T1 (Netherlands), PIC (Spain),
ASGC (Taiwan), RAL (UK) and BNL (USA) and in the Tier-2 facilities
worldwide.
\end{acknowledgments}

\providecommand{\href}[2]{#2}\begingroup\raggedright\endgroup

\onecolumngrid
\clearpage
\begin{flushleft}
{\Large The ATLAS Collaboration}

\bigskip

G.~Aad$^{\rm 48}$,
T.~Abajyan$^{\rm 21}$,
B.~Abbott$^{\rm 112}$,
J.~Abdallah$^{\rm 12}$,
S.~Abdel~Khalek$^{\rm 116}$,
O.~Abdinov$^{\rm 11}$,
R.~Aben$^{\rm 106}$,
B.~Abi$^{\rm 113}$,
M.~Abolins$^{\rm 89}$,
O.S.~AbouZeid$^{\rm 159}$
% ...
\bigskip
\\
% ...
$^{11}$ Institute of Physics, Azerbaijan Academy of Sciences, Baku, Azerbaijan\\
$^{12}$ Institut de F{\'\i}sica d'Altes Energies and Departament de F{\'\i}sica de la Universitat Aut{\`o}noma de Barcelona, Barcelona, Spain\\
% ...
$^{21}$ Physikalisches Institut, University of Bonn, Bonn, Germany\\
% ...
$^{48}$ Fakult{\"a}t f{\"u}r Mathematik und Physik, Albert-Ludwigs-Universit{\"a}t, Freiburg, Germany\\
% ...
$^{89}$ Department of Physics and Astronomy, Michigan State University, East Lansing MI, United States of America\\
% ...
$^{106}$ Nikhef National Institute for Subatomic Physics and University of Amsterdam, Amsterdam, Netherlands\\
% ...
$^{112}$ Homer L. Dodge Department of Physics and Astronomy, University of Oklahoma, Norman OK, United States of America\\
$^{113}$ Department of Physics, Oklahoma State University, Stillwater OK, United States of America\\
% ...
$^{116}$ LAL, Universit{\'e} Paris-Sud and CNRS/IN2P3, Orsay, France\\
% ...
$^{159}$ Department of Physics, University of Toronto, Toronto ON, Canada\\
% ...
$^{t}$ Also at Institut f{\"u}r Experimentalphysik, Universit{\"a}t Hamburg, Hamburg, Germany\\
% ...
\end{flushleft}


\begin{thebibliography}{10}
\bibitem{geometry}
ATLAS uses a right-handed coordinate system with its origin at the nominal
  interaction point~(IP) in the center of the detector and the $z$-axis along
  the beam pipe. The $x$-axis points from the IP to the center of the LHC ring,
  and the $y$-axis points upward. Cylindrical coordinates $(r,\phi)$ are used
  in the transverse plane, $\phi$ being the azimuthal angle around the beam
  pipe. The pseudorapidity is defined in terms of the polar angle $\theta$ as
  $\eta=-\ln\tan(\theta/2)$. The transverse energy $E_{\rm T}$ is defined as $E
  \sin{\theta}$, where $E$ is the energy associated to the calorimeter cell or
  energy cluster. Similarly, \pt\ is the momentum component transverse to the
  beam line.
\end{thebibliography}
\end{document}